**Simulation Study & At Home Diagnostic Tool for Early Detection of Parkinson's Disease**

Simoni Mishra, Mentor Matt Tivnan

**ABSTRACT**

Hypomimia is a condition in the early stages of the progression of Parkinson's disease that limits the movement of facial muscles, restricting accurate depiction of facial expressions. Also known as facial masking, this condition is an early symptom of Parkinson's disease, a neurodegenerative disorder that affects movement. To date, no specific test exists to diagnose the disease. Instead, doctors rely on the patient's medical history and symptoms to confirm the onslaught of the disease, delaying treatment. This study aims to develop a diagnostic tool for Parkinson's disease utilizing the Facial Action Coding System, a comprehensive system describing all facially discernible movement. In addition, this project generates image datasets by simulating faces or action unit sets for both Parkinson's patients and non-affected individuals through coding. Accordingly, the model is trained using supervised learning and neural networks. The model's efficiency is compared using Convolutional Neural Network and Vision Transformer models and later tested with a test program by passing an Parkinson's disease affected image as input. The future goal is to develop a self-determining mobile application (PDiagnosis) by utilizing the model to determine the probability of Parkinson's progression.

# INTRODUCTION

One of Parkinson's disease (PD) early signs is the gradual loss of facial mobility developing a "mask-like" appearance. This study investigates facial expressivity from simulated PD patients and Control populations and detects if the subject has progressed Parkinson's. In this work, 22,000 images (.png) files were generated using an OpenFACS API. Facial muscle movements were embedded in a neutral image using the facial action coding system. These images were generated for four facial expressions: happy, surprise, disgust, and neutral for both control and Parkinson's-affected subjects.

*Overview*

Clinical trials with human subjects are often not practical due to many limitations. However, virtual clinical trials and simulated subjects are slowly gaining popularity for clinical trials to evaluate and optimize many concepts and technologies. Due to the lack of datasets containing faces of Parkinson's patients, this study replaced human patients with simulated facial images of both Control and Parkinson's-affected patients. Facial Action Coding System (FACS) is a coded technique created by many researchers to understand facial muscle movements and expressions. Each muscle movement corresponds to a specific facial action unit. OpenFACS is an open-source, FACS based software that allows the simulation of facial expressions (Cuculo & D' Amelio). This system relies on Action Units for creating muscle movements. The OpenFACS API is open source free software available in Github. For this project, OpenFACS API is used to simulate Parkinson's images. Below is the list of facial action units defined by OpenFACs and the muscles associated with it.

| Action Unit | Description | Facial muscle |
|---|---|---|
| 1 | Inner Brow Raiser | Frontalis, pars medialis |
| 2 | Outer Brow Raiser | Frontalis, pars lateralis |
| 4 | Brow Lowerer | Depressor Glabellae, Depressor Supercilli, Currugator |
| 5 | Upper Lid Raiser | Levator palpebrae superioris |
| 6 | Cheek Raiser | Orbicularis oculi, pars orbitalis |
| 7 | Lid Tightener | Orbicularis oculi, pars palpebralis |

| 9 | Nose Wrinkler | Levator labii superioris alaeque nasi |
|---|---|---|
| 10 | Upper Lip Raiser | Levator Labii Superioris, Caput infraorbitalis |
| 12 | Lip Corner Puller | Zygomatic Major |
| 14 | Dimples | Buccinator |
| 15 | Lip Corner Depressor | Depressor anguli oris (Triangularis) |
| 17 | Chin Raiser | Mentalis |
| 20 | Lip Stretcher | Risorius |
| 23 | Lip Tightener | Orbicularis oris |
| 25 | Lips parter | Depressor Labii, Relaxation of Mentalis, Orbicularis Oris |
| 26 | Jaw Drop | Masetter; Temporal and Internal Pterygoid relaxed |

*Table 1. OpenFacs action units, colloquial terminology, and associated muscle groups (Ekman & Friesen, 1978)*

| **Facial Expressions** | **Action Units** |
|---|---|
| Disgust | (AU4+AU7+AU9) |
| Happy | (AU1+AU6+AU12) |
| Surprise | (AU1+AU2+AU4) |

*Table 2. OpenFacs Recommended Facial Expressions & Action Units (Ali et al.)*

## *Technology Used*

This analysis was conducted using Python 3.7, Tensorflow 2.5, Keras, Jupyter Notebook, Tensorboard, and NumPy: a scientific computing library that provides an efficient matrix and array calculations, Matplotlib: a plotting library used as an abstraction layer for creating models. The processing was done in an MSI laptop with Nvidia's GEforce RTX graphics card.

# METHODS

This study aims to build a model using facial expressions of the simulated images of Parkinson's and Control subjects to detect Parkinson disease. These simulated images are created by taking a neutral image and embedding the action units for various expressions into the images generating Control and Parkinson subjects. A Convolutional Neural Network model (CNN) model is trained using Images extracted from 11000 Parkinson and Control cases each. Later, a Vision Transformer model (Vit ) was created and trained with the same 22000 images to understand the difference between different models. It uses the following procedure to generate PD subjects and build the optimized model.

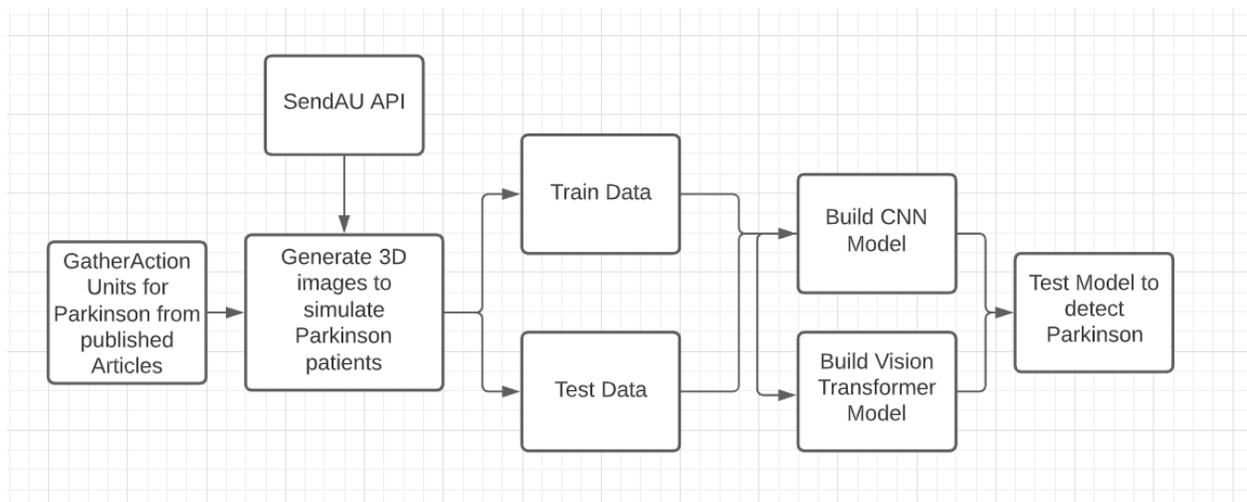

Figure 1. Overall Procedure

## Data Preparation

This study required data from the Parkinson's subjects which were not feasible due to lack of access to labs and human subjects during the Covid19 pandemic. One alternative solution was discovered to generate and simulate the patients using OpenFACS software. OpenFACS is an open-source API available freely in Github that generates the necessary png files with facial expressions. The building block of this API is based on action units. Action Units (AUs) are the fundamental actions of individual muscles or groups of muscles and external representations of muscle movements defined by their appearance on the face. This software required the Linux platform to operate. For this purpose, a system was configured by installing Ubuntu OS. The images were generated by calling the OpenFACS API's "sendAU" method. This method expects action units that comprise a set of facial muscles to create the image. The images were generated for the following facial expressions: Happy, Disgust, Surprise, and Neutral. However, the current OpenFACS API does not include any functionality to export or screen capture in any format, such as an image or video. After a few trial and error methods, an operating system command was discovered and issued to import the images from OpenFACS to save in the computer for training the models.

*Parkinson's Action Unit Detection*

For this study, a few research articles were used to understand the different muscle movements in the case of Parkinson's patients, especially in expressing various facial expressions. The following table from a research article gathers the action units variance and generates the simulated images of the Parkinson's patients in emotions like Smile, disgust, and surprise. Graphs and charts were studied to understand the variance of action units in Parkinson's cases and mapped to generate the Parkinson images. The table below describes the variance of facial Action Units between Control and Parkinson's cases(Ali et al., 2021).

| **Expression** | **Action Unit** | **Individuals with Parkinson's disease Mean (SD)** | **Individuals without Parkinson's disease Mean (SD** |
|---|---|---|---|
| Smiling Face | AU1 | 0.15(0.18) | 0.07(0.12) |
|  | AU6 | 0.17(0.15) | 0.25(0.25) |
|  | AU12 | 0.21(0.18) | 0.27(0.24) |
| Disgusted Face | AU4 | 0.19(0.20) | 0.26(0.31) |
|  | AU07 | 0.19(0.20) | 0.24(0.27) |
|  | AU09 | 0.04(0.06) | 0.04(0.07) |
| Surprise Face | AU01 | 0.28(0.28) | 0.27(0.32) |
|  | AU02 | 0.15(0.29) | 0.12(0.18) |
|  | AU04 | 0.31(0.37) | 0.40(0.43) |

*Table 3. Combinations of action units for the studied expressions (Derya & Kang)*

The following action units were taken by compiling the charts published in an article by Wu and Gonzalez.

| Emotions | Control Action Units (AU1,AU2,AU4,AU5,AU6,AU7,AU9,AU10,AU12,AU14,AU15,AU17,AU20,AU23,AU25,AU26,AU28,AU45) | Parkinson Action Units (AU1,AU2,AU4,AU5,AU6,AU7,AU9,AU10,AU12,AU14,AU15,AU17,AU20,AU23,AU25,AU26,AU28,AU45) |
|---|---|---|
| Disgust | [1,2,5,0,2,2,2,0,0,0,0,0,0,0,5,0,0,0] | [0,0,5,0,0,2,0,0,0,0,0,0,0,0,5,0,0,0] |
| Neutral | [0,0,1,0,0,2,0,0,0,0,0,0,0,5,0,0,0,2] | [0,0,5,0,0,2,2,0,0,0,0,0,1,0,5,0,0,0] |
| Happy | [5,0,0,0,5,0,0,0,5,0,0,0,0,0,0,0,0,0] | [3,0,0,0,3),0,0,0,3,0,0,0,0,0,0,0,0,0] |
| Surprise | [5,5,5,0,0,0,0,0,0,0,0,0,0,0,0,0,0,0] | [3,3,3),0,0,0,0,0,0,0,0,0,0,0,0,0,0,0] |

*Figure 2. Base parameters used for SendAU API for different expressions*

## *Data Generation*

The list of action units for each expression in Parkinson's was gathered from previous research articles, and the action units were computed using the variance from a prior study (Table3). The action units are shown above (Figure 2) were used as the base action units for the specific expression to generate simulated images for Parkinson's and Control cases.

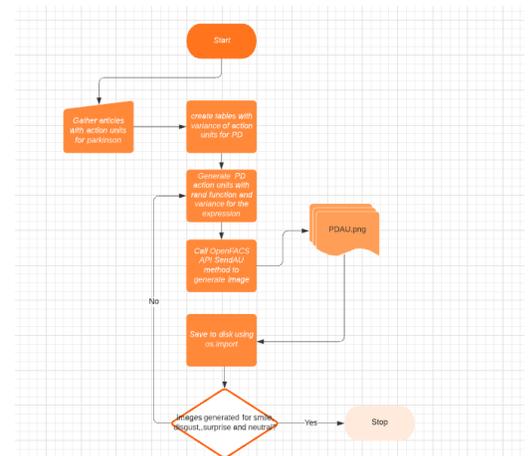

Numpy random function was used to apply a randomized variance value on speed and intensity of the muscle movements to create Parkinson's and Control cases. Action units for each emotion were passed as a JSON string to the "SendAU" method of OpenFACS API. Simulated images were imported and saved by providing the os.import command. For this study, 11,000 Parkinson's cases (images) and 11000 Control cases (images) were simulated. 2,750 images were generated for each expression. The code to generate the images by using the OpenFACS SendAPI is found in the appendix. Since currently, the OpenFACS API does not provide a way to save the images, the operating system "import" command was called to save the images to the disk. For all four expressions in both Parkinson's and Control images, the process was repeated.

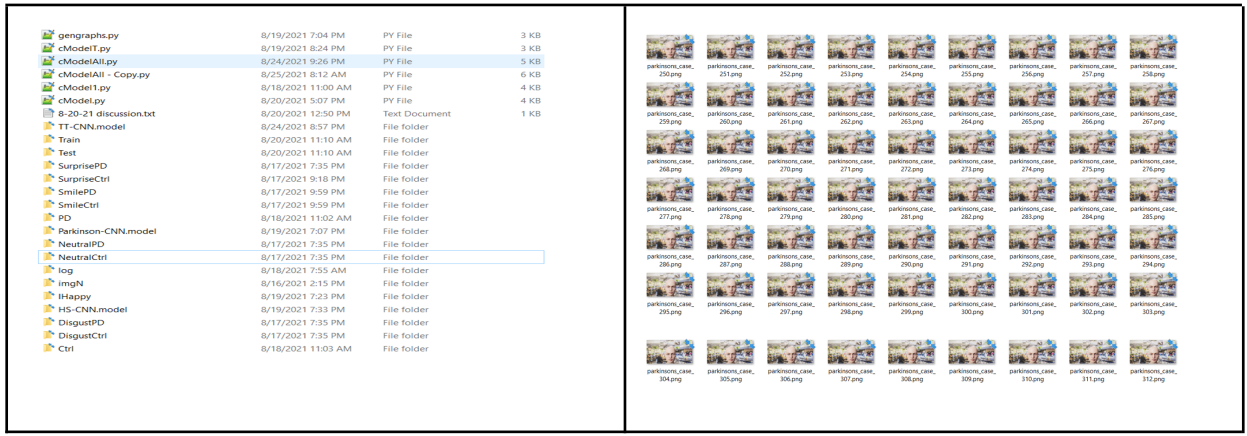

*Figure 4. Generated images CNN*

For this study, a total of 22000 simulated images were generated with varied emotions ranging from Disgust, Surprise, Neutral, and Happy (Ricciardi et al.,2017). A 75% and 25% ratio rule split the data between the training and test datasets.

|  | Parkinson images | Control images | Total |
|---|---|---|---|
| Train | 8800 | 8800 | 17600 |
| Test | 2200 | 2200 | 4400 |
| Total | 11000 | 11000 | 22000 |

| Subject Type | Disgust | Surprise | Neutral | Happy |
|---|---|---|---|---|
| Parkinson Train | 2200 | 2200 | 2200 | 2200 |
| Control Train | 2200 | 2200 | 2200 | 2200 |
| Parkinson Test | 550 | 550 | 550 | 550 |
| Control Test | 550 | 550 | 550 | 550 |

*Table 4. Number of simulated images created for Control & Parkinson's*

*Training & Test Data*

The table below shows a sample of the images generated for four different emotions for both simulated Parkinson's and Control subjects. As explained above, the images were generated by changing the facial action units. For this model, 17600 images are used to train the model, and 4400 images are used to test the model. There are two digits (0 to 1) or two classes to predict.

| | Parkinson's | Control |
|---|---|---|
| **Happy** | 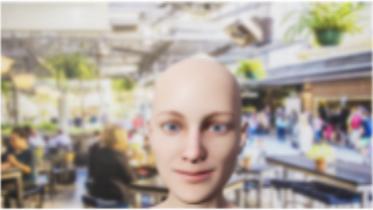 | 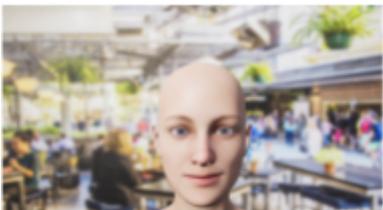 |
| **Surprise** | 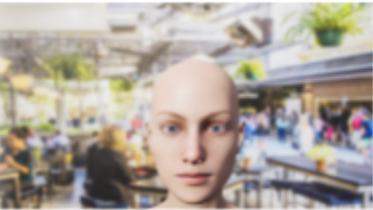 | 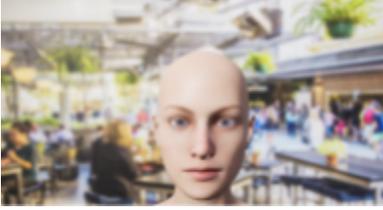 |
| **Disgust** | 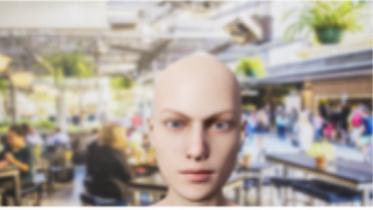 | 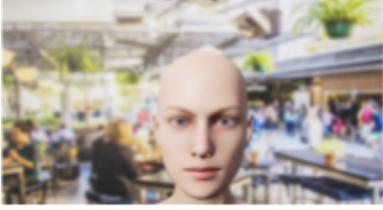 |
| **Neutral** | 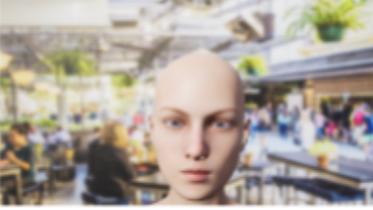 | 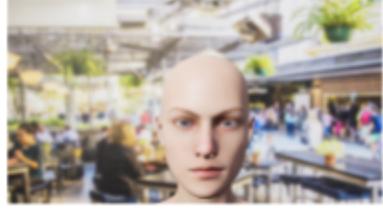 |

*Table 5: Showing sample images of expressions used in the study*

***Build Model (CNN & Vision Transformer (Vit))***

Initially, the model was built using CNN, which has specialized applications in image recognition (Singh, 2021). CNN model consists of three layers: convolutional layer, pooling layer, and fully connected layer. For this purpose, each image was resized into a 150 X 150-pixel square. The initial CNN model ran from 10 epochs to 100 epochs, measuring the accuracy, efficiency, and other issues like overfitting. However, at 50 epochs, the model was found to be effective. By studying the model summary and the charts from Tensorboard, it appeared that the pooling layer was losing a lot of information. So another attempt was made to experiment with the Vit model. However, Vit is a text-based task but works efficiently on input images as a sequence of image patches. This model divides an image into multiple square patches (Panteleyev, 2017). Each patch was flattened into a single vector and eventually helped the model to learn about each patch. The study used data augmentation and regularization to improve the model.

*Convolutional Neural Network (CNN)*

This multilayer deep learning model uses a neural network to pass every neuron from one layer to another. CNN provides promising results in image processing. A trained CNN has hidden layers whose neurons correspond to possible abstract representations over the input features. Since images are naturally non-linear, to increase the non-linearity, the rectifier (ReLU) function was applied. The image below shows the model summary of the CNN for this study. The dropout layer was applied to avoid overfitting, and the images were shuffled to reduce variance and provide better accuracy.

```
gpus1 [PhysicalDevice (name='/physical_device:GPU:0', device_type='GPU')]
Found 17600 images belonging to 2 classes.
Found 4400 images belonging to 2 classes.
Model: "sequential"
_________________________________________________________________
Layer (type)                 Output Shape              Param #
=================================================================
conv2d (Conv2D)              (None, 148, 148, 16)      448
_________________________________________________________________
max_pooling2d (MaxPooling2D) (None, 74, 74, 16)        0
_________________________________________________________________
conv2d_1 (Conv2D)            (None, 72, 72, 32)        4640
_________________________________________________________________
max_pooling2d_1 (MaxPooling2 (None, 36, 36, 32)        0
_________________________________________________________________
conv2d_2 (Conv2D)            (None, 34, 34, 64)        18496
_________________________________________________________________
max_pooling2d_2 (MaxPooling2 (None, 17, 17, 64)        0
_________________________________________________________________
flatten (Flatten)            (None, 18496)             0
_________________________________________________________________
dense (Dense)                (None, 512)               9470464
_________________________________________________________________
dense_1 (Dense)              (None, 1)                 513
=================================================================
Total params: 9,494,561
Trainable params: 9,494,561
Non-trainable params: 0
_________________________________________________________________
Epoch 1/50
880/880 [==============================] - 424s 474ms/step - loss: 0.5292 - accuracy: 0.7140 - val_loss: 0.1780 - val_accuracy: 0.9410
Epoch 2/50
880/880 [==============================] - 416s 473ms/step - loss: 0.1610 - accuracy: 0.9362 - val_loss: 0.1633 - val_accuracy: 0.9340
Epoch 3/50
880/880 [==============================] - 460s 523ms/step - loss: 0.1265 - accuracy: 0.9495 - val_loss: 0.0850 - val_accuracy: 0.9660
Epoch 4/50
880/880 [==============================] - 417s 474ms/step - loss: 0.1087 - accuracy: 0.9576 - val_loss: 0.1339 - val_accuracy: 0.9480
Epoch 5/50
880/880 [==============================] - 416s 472ms/step - loss: 0.1002
```

*Figure 5. Model summary of the CNN model*

In the first few runs, the accuracy of the 1st epoch was at 71 % and increased later as the epoch increased. Later, the dropout layer was added to generalize and reduce overfitting. With dropout, the neurons from the current layer randomly disconnect from the successive layers so that it relies on the existing connections reducing the probability of overfitting. Here are the details from the summary:

Flatten takes 2D matrix and creates 1D output, giving 17 X 17 X 64 = 18496 pixels

The First dense layer is 512 nodes - A weight per 18496 pixels which is connected to the nodes 512 X 18496 = 9470464

Second dense layer is a weight per connection to node plus bias per node. Total = 512 +1 = 513

Putting this all together we get, Conv2D: 448, Conv2D_1: 4640, Conv2D_2: 18496, Dense: 9470464, Dense_1: 513, Total = 448+ 4640+ 18496 + 9470464 + 513 = 9494561

```
Total params: 9,494,561
Trainable params: 9,494,561
Non-trainable params: 0
_________________________________________________________________
Epoch 1/50
880/880 [==============================] - 424s 474ms/step - loss: 0.5292 - accuracy: 0.7140 - val_loss: 0.1780 - val_accurac
y: 0.9410
Epoch 2/50
880/880 [==============================] - 416s 473ms/step - loss: 0.1610 - accuracy: 0.9362 - val_loss: 0.1633 - val_accurac
y: 0.9340
Epoch 3/50
880/880 [==============================] - 460s 523ms/step - loss: 0.1265 - accuracy: 0.9495 - val_loss: 0.0850 - val_accurac
y: 0.9660
Epoch 4/50
880/880 [==============================] - 417s 474ms/step - loss: 0.1087 - accuracy: 0.9576 - val_loss: 0.1339 - val_accurac
y: 0.9480
```

*Figure 6. Model Building*

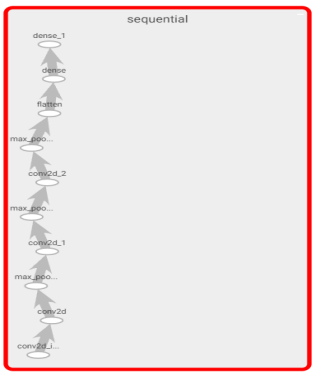

Here is a conceptual graph of the neural network model in Tensorboard, which shows different network layers. The section below utilizes Tensorboard to understand the inner workings, especially the numerous biases and weights associated with each layer. The picture below shows the histogram of the hyperparameters in this model, which ran for 50 epochs.

*Figure 7. Sequential Model (CNN Conceptual view)*

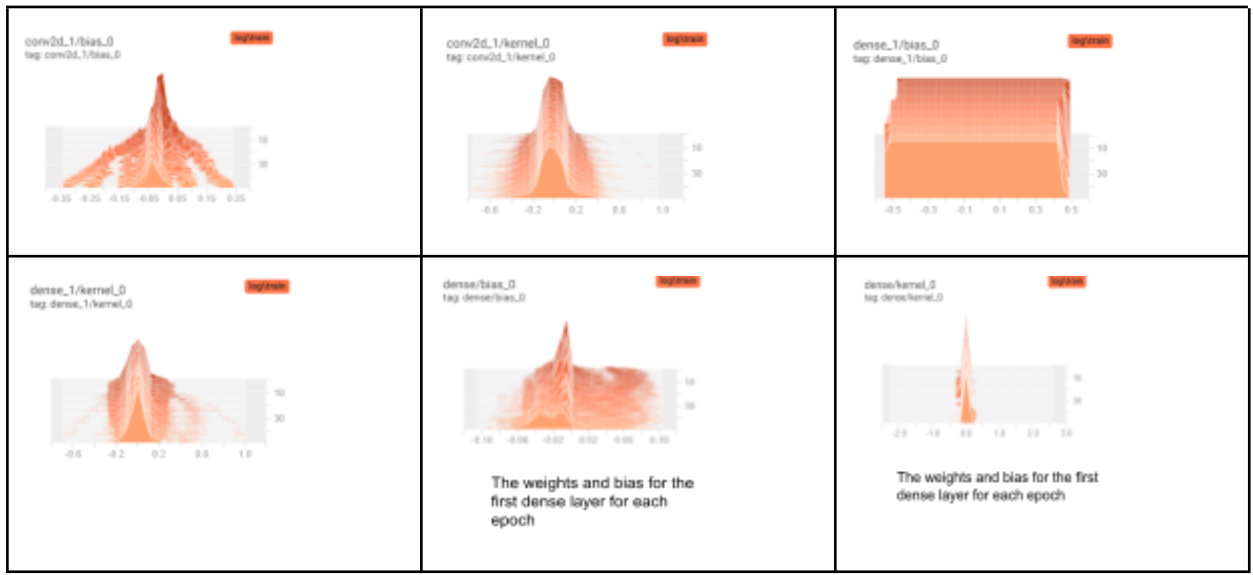

*Figure 8. Histograms of CNN from Tensorboard*

### Vision Transformer (Vit)

The Vit model for image classification uses a Transformer-like architecture. For this, an image was split into fixed-size patches, each of which was then linearly embedded, position embeddings were added, and the resulting sequence of vectors was fed to a standard transformer encoder to pre-train datasets. For this analysis, in the first step of building the model, the image was divided into sequences of image patches divided into 6 X 6 dimensions. Later, it was increased to 10 X 10 dimensions. The transformer encoder module comprises a multi-head attention and perception layer. After applying normalization layers, an MLP (multilayer perceptron) head was added to the transformer to result in the output class. The model was debugged using Tensorboard charts and optimized accordingly. Regularizations were applied to normalize and improve the accuracy.

Model Summary:

```
Model: "model"
__________________________________________________________________________________________________
Layer (type)                    Output Shape         Param #     Connected to
==================================================================================================
input_1 (InputLayer)            [(None, 100, 100, 3) 0
__________________________________________________________________________________________________
data_augmentation (Sequential)  (None, 72, 72, 3)    7           input_1[0][0]
__________________________________________________________________________________________________
patches (Patches)               (None, None, 108)    0           data_augmentation[0][0]
__________________________________________________________________________________________________
patch_encoder (PatchEncoder)    (None, 144, 64)      16192       patches[0][0]
__________________________________________________________________________________________________
layer_normalization (LayerNorma (None, 144, 64)      128         patch_encoder[0][0]
__________________________________________________________________________________________________
multi_head_attention (MultiHead (None, 144, 64)      66368       layer_normalization[0][0]
                                                                 layer_normalization[0][0]
__________________________________________________________________________________________________
add (Add)                       (None, 144, 64)      0           multi_head_attention[0][0]
                                                                 patch_encoder[0][0]
__________________________________________________________________________________________________
layer_normalization_1 (LayerNor (None, 144, 64)      128         add[0][0]
__________________________________________________________________________________________________
dense_1 (Dense)                 (None, 144, 128)     8320        layer_normalization_1[0][0]
```

*Figure 9. model Summary for the VisionTransformer*

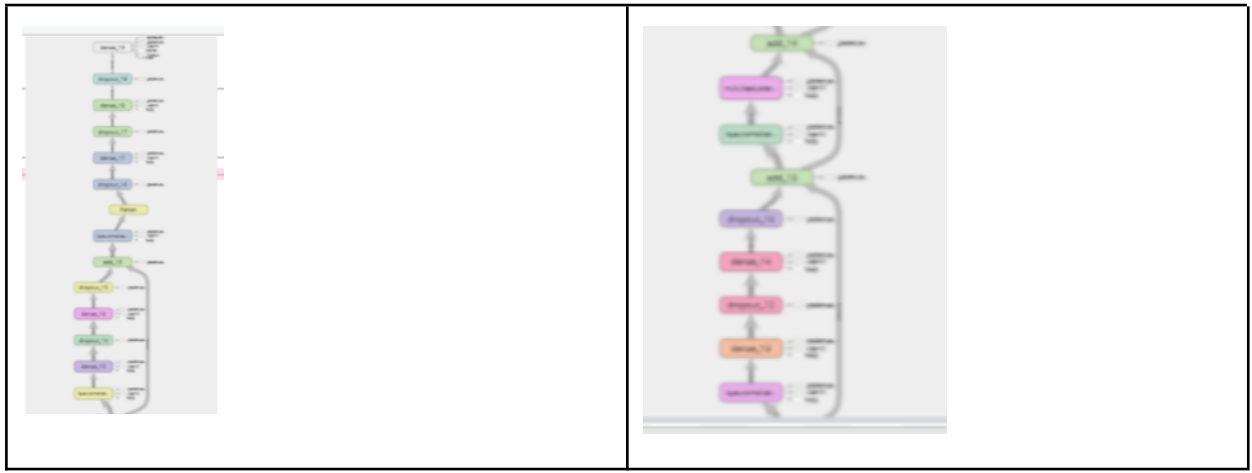

*Figure 10. Vision Transformer architecture*

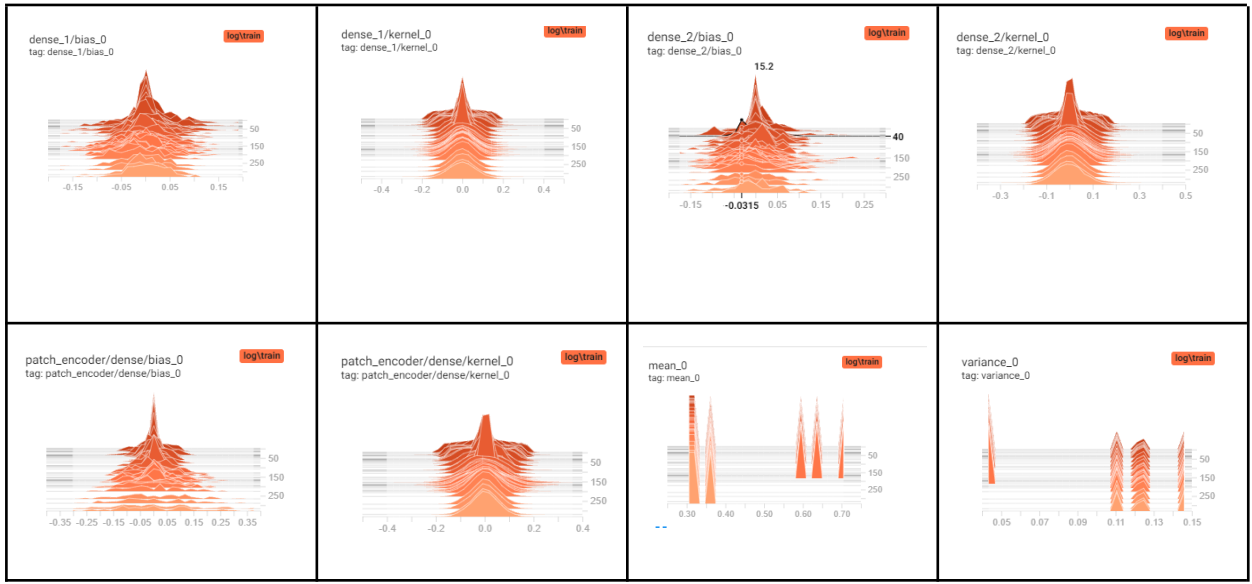

*Figure 11. Histograms of Vision transformer from Tensorboard*

## RESULTS

Tensorboard generates the histogram and shows the changes in loss and accuracy after every epoch when the data was passed through a CNN model.

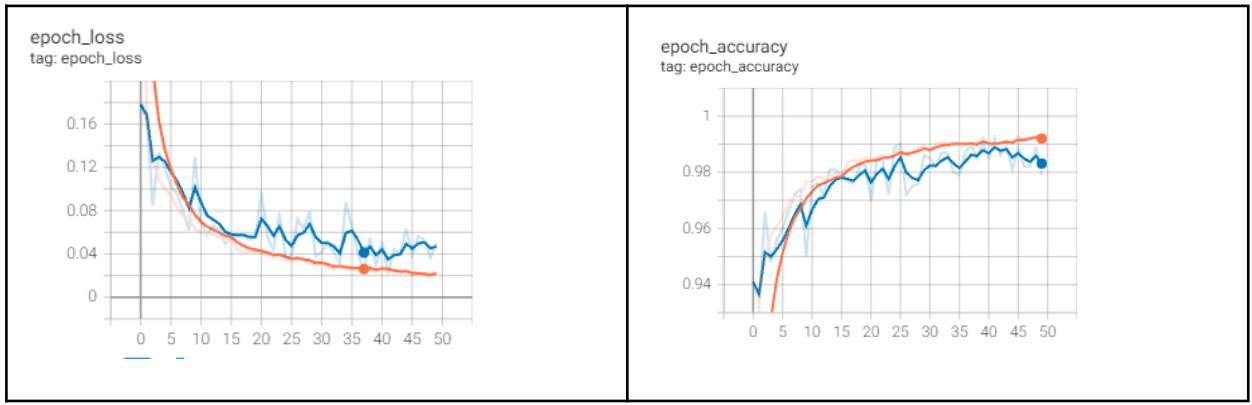

*Figure 12. CNN model: Training & Validation Loss vs. Accuracy*

As the figure shows, in the CNN model, the validation (orange line) loss reduced as the epoch increased; alternatively, in the chart to the right, it shows that the accuracy increased as the epoch increased.

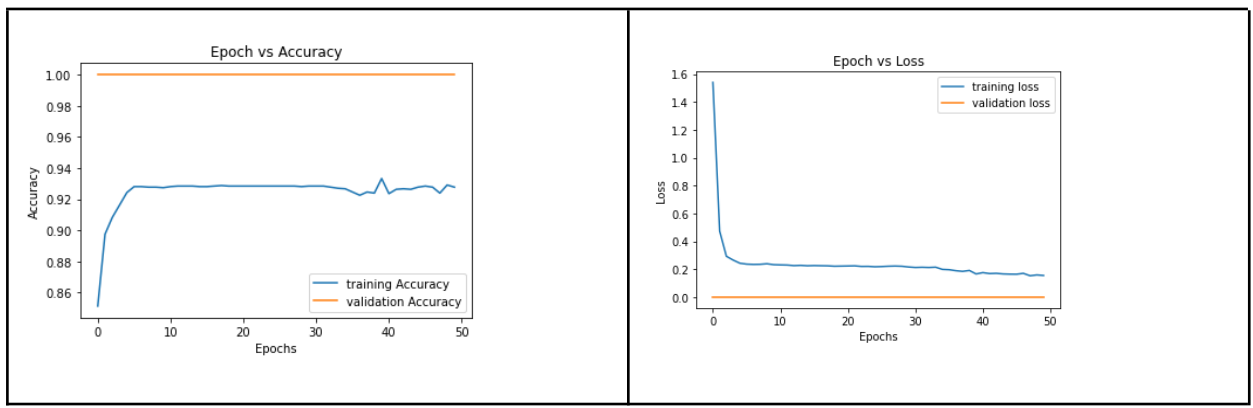

*Figure 13. Transformer model: Training & Validation Loss vs. Accuracy*

The Vit model was trained using the same number of simulated Parkinson's and Control subjects as the CNN model. The small number of images used for the study was a limitation for this model, and a few issues were observed, which were improved through optimization and regularization. As shown above in Figure 13, after a few epochs, the dataset was underfitting which was difficult to optimize. A test program was created to test the model by inputting the Parkinson's and Non-Parkinson simulated images.

| Type | CNN | Vit |
|---|---|---|
| Total Parameters | 9,494,561 | 21,759,019 |
| Trainable Parameters | 9,494,561 | 21,759,012 |
| Non-Trainable Parameters | 0 | 7 |
| Prediction Accuracy | 100% successful prediction | 60 % successful prediction |
| Processing time to generate models (GPU based) | 90 minutes for 50 epochs | 10 minutes for 50 epochs |
| Accuracy | 98% | 56% |

*Table 5. CNN vs. Transformer model*

*Test A Model*

As a part of this study, both models were tested using a test program. It was tested by passing the appropriate image as the input parameter to the test program. The model returns "1" in the case of Parkinson's and "0" in the Non-Parkinson case. The code to test the model is provided in the appendix. Both the models (CNN & Vit) were tested using a Test program. Though the expectation was to see a better result in the Vit model, CNN provided 98% prediction accuracy with 20 test cases in this study. In contrast, Vit achieved a 60% prediction accuracy with 14 test cases passed out of 20 total test cases.

**DISCUSSION**

This study used action units to generate virtual Parkinson's subjects to conduct experiments. The muscle movements' variance was calculated using the random function to show that micro-expressions are important biomarkers for Parkinson's Disease. The study was conducted using the CNN model, and later the Vit model was used to observe the difference. The first set of analyses looks at the difference in the distribution of features. The transformer mechanism enhances the focused parts of the data and fades out the rest. CNN does not extract the relative position of the feature rather a combination of features. Though CNN is very good at feature extraction, filters for large receptive fields are required to track long-range features.

Furthermore, this reduces the efficiency of the model. Initially, the CNN model suffered some level of overfitting, and later the issue was fixed by bringing regularization. After 50 epochs, the Vit model achieved around 56% accuracy on the test data. The CNN classifier achieved about 98% accuracy, and the model was tested using a test program. To improve the quality of the Vit model, different parameters such as increasing the number of epochs, increasing the number of transformer layers, increasing patch size, and resizing the input images were added. After training the Vit model to 400 epochs, the accuracy increased to 80%. After working with this model, it was noticed that it generalizes poorly in small datasets. Based on the recommendations from a few other articles, the Vit model is expected to perform much better with a large high-resolution dataset.

As mentioned earlier, the test result provides a way to detect Parkinson's from everyday activities. These findings provide evidence that the early detection of Parkinson's through images and facial expressions could soon become a reality using mobile applications.

OpenFACS API allows the generation of 3D images (animated). However, due to a technical constraint, 3D images couldn't be saved for analysis; instead, images were generated. The future effort will include 3D images.

*Application*

Currently, an effort is underway to create a mobile application using this model to read the user's picture from a mobile device and diagnose if the user has Parkinson's disease. This mobile application uses TensorFlow Lite to run the application on a mobile device.

**CONCLUSION**

This proof-of-principle demonstrated two approaches. It used a facial action coding system to generate Parkinson's subjects for study and later created a model to detect Parkinson's from the hypomimia cases. The goal of this study is to use this model in a mobile application called PDiagnosis, which can be used as an at-home Parkinson's detection tool to detect the progression of Parkinson's from the user's daily activities like when taking pictures or by integrating with Facial recognition software (Patil, 2021). The results of this study open new avenues for future research and may serve as a source of hypothesis generation for future researchers. The approach of generating simulated images using action units can be used for other studies without using human subjects. PDiagonsis can be extended to use human subjects and use a high volume of real images to improve accuracy.

**APPENDIX**

*Code and Result of Simulated Subject*

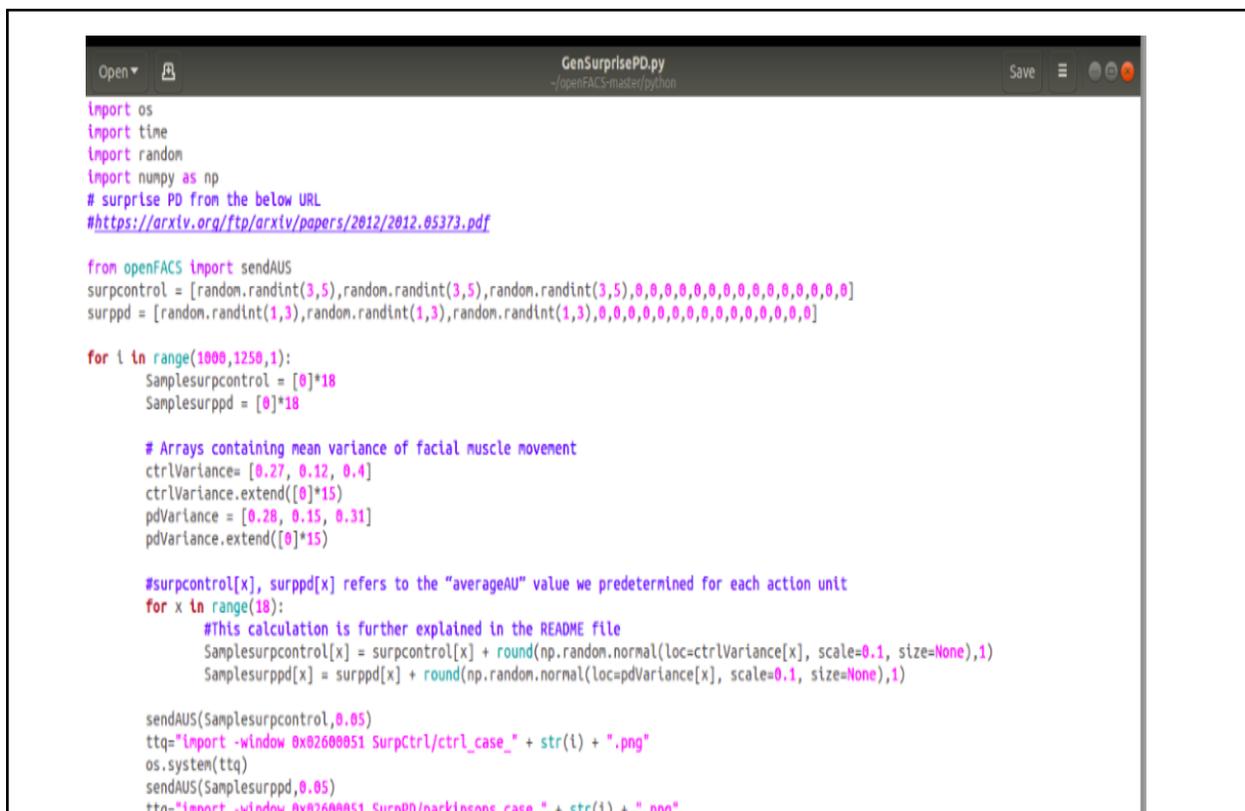

*Figure 14. Code to generate images*

```
In [2]: import os
        import tensorflow as tf
        import numpy as np

        from keras.preprocessing import image

        # predict using the model

        #ttq="C:\\poly1\\train\\ctrl_case_7.png"
        ttq="C:\\npoly\\train\\ctrl_case_1999.png"
        #ttq="C:\\npoly\\train\\parkinsons_case_1999.png"

        img = image.load_img(ttq, target_size=(150, 150))
        x = image.img_to_array(img)
        x = np.expand_dims(x, axis=0)

        x_test = np.zeros([1,20,0,4])

        model = tf.keras.models.load_model("c:\\npoly\\rpsPD.h5")
        #model = tf.keras.models.load_model("c:\\polyTRANS\\rpsPD100421.MODEL")

        images = np.vstack([x])
        classes = model.predict(images, batch_size=10)

        print(classes)

        [[0.]]
```

*Figure 15. Code to Test model - 0 results for Non-PD*

***Test CNN Model***

```
In [1]: import os
        import tensorflow as tf
        import numpy as np

        from keras.preprocessing import image

        # predict using the model

        #ttq="C:\\poly1\\train\\ctrl_case_7.png"
        #ttq="C:\\npoly\\train\\ctrl_case_1999.png"
        ttq="C:\\npoly\\train\\parkinsons_case_1999.png"

        img = image.load_img(ttq, target_size=(150, 150))
        x = image.img_to_array(img)
        x = np.expand_dims(x, axis=0)

        x_test = np.zeros([1,20,0,4])

        model = tf.keras.models.load_model("c:\\npoly\\rpsPD.h5")
        #model = tf.keras.models.load_model("c:\\polyTRANS\\rpsPD100421.MODEL")

        images = np.vstack([x])
        classes = model.predict(images, batch_size=10)

        print(classes)
```

```
WARNING:tensorflow:From C:\Users\bimal\AppData\Roaming\Python\Python37\site-packages\tensorflow\python\ops\array_ops.py:5075: c
alling gather (from tensorflow.python.ops.array_ops) with validate_indices is deprecated and will be removed in a future versio
n.
Instructions for updating:
The `validate_indices` argument has no effect. Indices are always validated on CPU and never validated on GPU.
[[1.]]
```

*Figure 16. Code to Test the CNN model - result 1 for PD*

*Source Code*

*Figure 17. Code to train for Vit model*

*Figure 18. Code to train CNN model*